\documentclass[11pt]{article}
\usepackage{aas_macros,amsmath,amssymb,comment,cite,esint,graphicx,mathtools,diagbox}
\usepackage{bm}
\usepackage[margin=.8in,letterpaper]{geometry}
\usepackage[colorlinks=true]{hyperref}
\usepackage[affil-it]{authblk}
\usepackage{subcaption}
\usepackage[utf8]{inputenc}
\usepackage{mathrsfs}
\usepackage{appendix}
\usepackage{amssymb}
\usepackage{float}                  % 图片浮动位置,强制H
\usepackage{color}
\usepackage{cite}
\usepackage{hyperref}
\hypersetup{pageanchor=false}
\usepackage{indentfirst}
\usepackage{url}
\usepackage{xfrac}
\usepackage{caption}
\usepackage[numbers,square,comma,sort&compress,merge]{natbib}
\usepackage{esint}
\usepackage{overpic}
\usepackage{graphicx}
\usepackage{epsf,amsmath,bbold,amsfonts,stmaryrd}
\usepackage{textcomp}
\usepackage{ulem}
\usepackage{tikz}
\usepackage{multirow}
\numberwithin{equation}{section}
\let\originalleft\left
\let\originalright\right
\renewcommand{\left}{\mathopen{}\mathclose\bgroup\originalleft}
\renewcommand{\right}{\aftergroup\egroup\originalright}
\def\bea{\begin{eqnarray}}
\def\eea{\end{eqnarray}}
\def\nn{\nonumber}

\usepackage{tensor}
\usepackage{physics}

\newcolumntype{P}[1]{>{\Centering\hspace{0pt}}p{#1}}
\newcolumntype{Z}{>{\centering\arraybackslash}X} %Z单元格居中

\newcommand{\df}{\mathrm{d}}   %微分符号
\begin{document}
\title{\bf Geodesics and Shadows in the Kerr–Bertotti–Robinson Black Hole Spacetime }
	
\author{Xinyu Wang$^{1,2}$, Yehui Hou$^3$, Xi Wan$^{4}$, Minyong Guo$^{1,2\ast}$, Bin Chen$^{4,5\ast}$}
\date{}
	
\maketitle
\vspace{-15mm}

\begin{center}
{\it
$^1$ School of physics and astronomy, Beijing Normal University,
Beijing 100875, P. R. China\\\vspace{2mm}

$^2$ Key Laboratory of Multiscale Spin Physics (Ministry of Education), Beijing Normal University, Beijing 100875, China\\\vspace{2mm}

$^3$ Tsung-Dao Lee Institute, Shanghai Jiao-Tong University, Shanghai, 201210, P. R. China\\\vspace{2mm}

$^4$ Institute of Fundamental Physics and Quantum Technology, \\
\& School of Physical Science and Technology,\\Ningbo University, Ningbo, Zhejiang 315211, China\\\vspace{2mm}

$^5$ School of Physics, Peking University, \& Center for High Energy Physics, \\No.5 Yiheyuan Rd, Beijing 100871, P.R. China\\\vspace{2mm}

}
\end{center}

\vspace{8mm}
\begin{abstract}
In this work, we investigate geodesics and black hole shadows in the Kerr–Bertotti–Robinson spacetime. We show that the equations of motion for null geodesics are separable and admit analytical treatment, whereas timelike geodesics are generally non-separable. Approximate analytical expressions for the photon sphere and the innermost stable circular orbit are derived via perturbative expansions in the magnetic field strength.
We further explore the black hole shadow using both numerical and analytical methods, examining the effects of the magnetic field, the observer’s inclination angle and radial position. Deviations from the standard Kerr shadow are quantified, and a physical interpretation is provided by introducing asymptotic regimes defined relative to the magnetic field strength.
\end{abstract}

\vfill{\footnotesize $\ast$ Corresponding authors: minyongguo@bnu.edu.cn, chenbin1@nbu.edu.cn}

\maketitle

\newpage
\baselineskip 18pt
\section{Introduction}\label{sec1}

Black holes in the universe do not exist in isolation; growing observational evidence indicates that magnetic fields can permeate their surrounding environments. In astrophysical settings, these magnetic fields can reach extreme strengths. Notably, the strongest magnetic field observed near a black hole—on the order of $10^{14}$ Gauss—originates from the nearby magnetar SGR J1745–29, located close to Sgr A* \cite{Eatough:2013nva,Kennea:2013dfa,Olausen:2013bpa}. 
The presence of magnetic fields near black holes gives rise to a range of rich physical phenomena, including synchrotron radiation from accelerated electrons, magnetic reconnection, the formation of magnetized accretion disks \cite{Abramowicz:2011xu}, and the launching of relativistic jets \cite{1977MNRAS.179..433B,1982MNRAS.199..883B}.

When the magnetic field is relatively weak, its backreaction on the spacetime geometry can be safely neglected. In this regime, Wald identified a solution to the source-free Maxwell equations that describes a weak, uniform magnetic field in the Kerr spacetime \cite{Wald:1974np}. However, as the magnetic field strength increases, its self-gravitational effects can no longer be ignored. In response to this, Ernst derived exact solutions to the coupled Einstein–Maxwell equations, representing Schwarzschild and Kerr black holes immersed in the Melvin universe \cite{Ernst:1976mzr, Ernst:1976bsr}. These configurations—commonly known as the Schwarzschild–Melvin and Kerr–Melvin black holes—incorporate a vertical magnetic field directly encoded in the spacetime metric.
More recently, Podolský and Ovcharenko introduced a new class of exact solutions describing a Kerr black hole embedded in the Bertotti–Robinson (BR) universe, permeated by a uniform Maxwell field \cite{Podolsky:2025tle}. These spacetimes are known as Kerr–Bertotti–Robinson (KBR) black holes.

In contrast to Kerr–Melvin black holes, which are of Petrov type I \cite{Barrientos:2024pkt}, KBR black holes belong to the Petrov type D class \cite{petrov1954classification}. Notably, they feature bounded ergoregions, and their electromagnetic fields remain asymptotically finite and uniform. In principle, test particles can escape to infinity. These properties make KBR black holes a more physically realistic model of a black hole immersed in an external electromagnetic field.
The Schwarzschild–Melvin and Kerr–Melvin solutions have been extensively studied, covering topics such as geodesic structure \cite{Stuchlik:1999mro, 1983JPhA...16...99D, 1984NCimB..79...76E}, the dynamics of charged particles \cite{1979JPhA...12..215D, 2019EPJP..134...96L, Shaymatov:2022enf}, black hole shadows \cite{Junior:2021dyw, Wang:2021ara}, photon rings, and image formation from luminous accretion disks \cite{Hou:2022eev}. It is therefore natural to explore these phenomena in the KBR spacetime.
It is also worth noting that magnetic reconnection in the presence of strong magnetic fields can be significant \cite{Fan:2024fcy, Fan:2024rsa}, potentially triggering an astrophysically realistic Penrose process within the ergosphere \cite{Koide:2008xr, Comisso:2020ykg, Chen:2024ggq, Zhang:2024ptp, Shen:2024sdr}, as recently explored for the KBR black hole in \cite{Zeng:2025olq}.

Moreover, given the growing importance of black hole shadow and image studies \cite{Gralla:2019xty, Junior:2021svb, Chang:2021ngy, Guo:2022nto, He:2022opa, Zhu:2022shb, Hosseinifar:2024wwe, Luo:2024avl, Liu:2024soc, Zeng:2025nmu, Zeng:2025kqw, Kuang:2024ugn, Guerrero:2022msp, Zhu:2022amy, Hu:2023pyd, Ye:2023qks,  Heydari-Fard:2023kgf, Meng:2023uws, Darvishi:2024ndu, Guo:2024mij, Li:2025awg, Liang:2025bbn, Meng:2025ivb, Huang:2025xqd, Li:2025jfq, Liu:2025lwj, Wang:2025ihg, Xu:2025iwg, Xu:2025jvk, Guo:2025yin,Chen:2024oyv,Jafarzade:2025byr,Wang:2025fmz,Lee:2021sws,Tsukamoto:2014tja}—particularly in light of the Event Horizon Telescope’s groundbreaking images of the supermassive black holes at the centers of the M87 galaxy and the Milky Way \cite{EventHorizonTelescope:2019dse, EventHorizonTelescope:2022wkp, EventHorizonTelescope:2021bee, EventHorizonTelescope:2021srq}—the KBR black hole, as a novel class of exact solutions, merits dedicated investigation.
In this work, we focus on the analysis of geodesics and the shadow of the KBR black hole, thereby laying the foundation for more comprehensive investigations of its observational features. We show that null geodesics remain analytically tractable due to the separability of the Hamilton–Jacobi equation, whereas timelike geodesics are generally non-separable.
We derive approximate analytic expressions for the photon sphere and the innermost stable circular orbit (ISCO), expanded perturbatively in the magnetic field strength. The black hole shadow is then computed using both analytical approximations and numerical ray-tracing techniques. To quantify deviations from the Kerr shadow, we introduce a deformation parameter and analyze its dependence on the magnetic field strength, the observer’s inclination angle, and radial position.

The remaining sections of this paper are structured as follows. In Sec. \ref{sec2}, we review the KBR black hole spacetime and discuss its basic properties. Sec. \ref{sec3} presents a comprehensive and systematic investigation of the null and timelike geodesics within the KBR black hole spacetime. In Sec. \ref{sec4}, we explore the shadow cast by the KBR black hole and quantitatively examine its deviations from, and distinctions relative to, that of the Kerr black hole. Finally, in Sec. \ref{sec5}, we provide a summary and outlook. In this work, we have set the fundamental constants $c$, $G$ to unity, and we will work in the convention $(-,+,+,+)$.

\section{The KBR Spacetime}\label{sec2}

In this section, we aim to provide a review of the KBR black hole, which was recently discovered in \cite{Podolsky:2025tle}. Its line element can be expressed as  
\begin{align}
    \df s^2 = \dfrac{1}{\Omega^2} \Big[-\frac{Q}{\Sigma}
    \big(\df t - a \sin^2\theta\, \df\varphi\big)^2
    + \frac{\Sigma}{Q}\, \df r^2
    + \frac{\Sigma}{P}\, \df\theta^2
    + \frac{P}{\Sigma} \sin^2\theta\,
    \big(a\, \df t - (r^2 + a^2)\, \df\varphi\big)^2 \,\Big]\,, \label{KBR}
\end{align}  
where the metric functions are given by  
\begin{align}
\Sigma   &= r^2 + a^2 \cos^2\theta\,, \label{rho2}\\[0mm]
P &= 1 + B^2 \left( m^2\,\dfrac{I_2}{I_1^2} - a^2 \right) \cos^2\theta\,, \label{tilde_P}\\
Q &= \left(1 + B^2 r^2 \right) \Delta\,, \label{math-Q}\\[2mm]
\Omega^2 &= \left(1 + B^2 r^2 \right) - B^2 \Delta \cos^2\theta\,, \label{Omega}\\
\Delta &= \left(1 - B^2 m^2 \dfrac{I_2}{I_1^2}\right) r^2 - 2m\,\dfrac{I_2}{I_1}\, r + a^2\,, \label{Delta}
\end{align}  
with  
\begin{align}
    I_1 = 1 - \tfrac{1}{2} B^2 a^2\,, \qquad
    I_2 = 1 - B^2 a^2\,. \label{I1I2}
\end{align}  
Here, $m$ denotes the mass parameter of the central black hole, $a$ is the spin parameter, and $B$ characterizes the strength of the electromagnetic field. 
The spacetime given by Eq.~\eqref{KBR} satisfies the Einstein–Maxwell equations, where the 1-form gauge potential is given by
\begin{align}
A_{\mu} \, \df x^{\mu} = \dfrac{{\rm e}^{\mathrm{i} \gamma}}{2B} \left[ \partial_r\Omega \, \dfrac{a \, \df t - (r^2 + a^2) \, \df\varphi}{r + \mathrm{i} a \cos\theta} + \dfrac{\mathrm{i} \, \partial_\theta\Omega}{\sin\theta} \, \dfrac{\df t - a \sin^2\theta \, \df\varphi}{r + \mathrm{i} a \cos\theta} + (\Omega - 1) \, \df\varphi \right]\,,  
\label{Amu}
\end{align}  
which vanishes in the limit $B = 0$. Here, $\mathrm{i}$ denotes the imaginary unit, and it is the real part of Eq.~\eqref{Amu} that corresponds to the physically meaningful electromagnetic potential. Furthermore, $\gamma$ represents an arbitrary phase parameter that governs the relative contributions of the electric and magnetic components. In particular, setting $\gamma = 0$ yields a purely magnetic configuration, whereas $\gamma = \pi / 2$ corresponds to a purely electric field. Crucially, the choice of $\gamma$ has no influence on the geometry of the spacetime.
Furthermore, the metric belongs to Petrov type D—distinct from the type I structure of the Kerr–Melvin solution—and features a source-free electromagnetic field that is neither null nor aligned with the principal null directions \cite{Podolsky:2025tle}.

It is worth noting that, considering the regularity of the KBR metric at $\theta = 0$ and $\theta = \pi$, we require that  
\bea
2\pi &= \lim_{\theta \to 0} \frac{\text{circumference}}{\text{radius}} = \int_0^{\varphi_{\text{max}}} \sqrt{g_{\varphi\varphi}} \, d\varphi = \varphi_{\text{max}} P(0)\,, \nn \\
2\pi &= \lim_{\theta \to \pi} \frac{\text{circumference}}{\text{radius}} = \int_0^{\varphi_{\text{max}}} \sqrt{g_{\varphi\varphi}} \, d\varphi = \varphi_{\text{max}} P(\pi)\,.
\eea  
Thus, for the sake of convention, we define $\phi = C^{-1}\varphi$, where 
\[
C = P^{-1}(0) = P^{-1}(\pi) = \left[1 + B^2 \left(m^2 \frac{I_2}{I_1^2} - a^2 \right)\right]^{-1}\,.
\]
so that $\phi \in [0, 2\pi)$. Then the metric in Eq. (\ref{KBR}) can be rewritten as 
\begin{align}\label{KBR02}
    \df s^2 = \dfrac{1}{\Omega^2} \Big[-\frac{Q}{\Sigma}
    \big(\df t - a \sin^2\theta\, C\df\phi\big)^2
    + \frac{\Sigma}{Q}\, \df r^2
    + \frac{\Sigma}{P}\, \df\theta^2
    + \frac{P}{\Sigma} \sin^2\theta\,
    \big(a\, \df t - (r^2 + a^2)\, C\df\phi\big)^2 \,\Big]\,.
\end{align}  
From $Q=$ corresponding to $\Delta=0$ yields up to two solutions:  
\begin{equation}
r_\pm = \frac{m I_2 \pm \sqrt{m^2 I_2 - a^2 I_1^2 }}{I_1^2 - B^2 m^2 I_2}I_1\,,
\end{equation}  
which correspond to the locations of the outer and inner horizons of the black hole. When the expression under the square root equals zero, it corresponds to the extreme situation, characterized by the condition  
\[
B^2 =\frac{2}{a^4} \left( m - \sqrt{m^2 - a^2} \right) \sqrt{m^2 - a^2}\,.
\]
From this expression, we observe that when the magnetic field is nonzero, the maximum value of $a$ is less than $m$. This implies that, in the presence of a non-vanishing magnetic field, the spin parameter $a$ of an extremal KBR black hole is strictly less than its mass parameter.
In the absence of the Maxwell field, namely when $B = 0$, the metric reduces to the Kerr solution expressed in Boyer–Lindquist coordinates. In the limiting case $m \to 0$, the geometry approaches the BR metric \cite{Al-Badawi:2004agm}, which exhibits the symmetry of a direct product spacetime, $\text{AdS}_2 \times S^2$. When both $B$ and $m$ vanish, the metric degenerates into flat Minkowski spacetime. If only the rotation parameter $a$ is set to zero, the spacetime reduces to the Schwarzschild–BR geometry. The parameters $B$ and $a$ possess dimensions of inverse length and length, respectively:
\[
[B] = [a^{-1}] = [m^{-1}] = [r^{-1}].
\]
In the present analysis, we concentrate on the regime where $m$ is nonzero. Upon translating into Gaussian units, the magnetic field strength is given by \cite{Hou:2022eev}
\[
B_{\text{Gauss}} = \frac{c^4}{G^{3/2} m} (Bm) = 2.36 \times 10^{19} \left( \frac{M_\odot}{m} \right) (Bm),
\]
where $M_\odot$ denotes the solar mass; $B$ denotes the magnetic field strength in natural units (with $G = c = 1$), and $B_{\text{Gauss}}$ represents its corresponding value in Gauss. For a supermassive black hole with a mass on the order of $m \sim 10^9 M_\odot$, a dimensionless field strength of $Bm = 0.01$ translates to a physical magnetic field of approximately $B_{\text{Gauss}} \sim 10^8 \ \text{Gauss}$, a magnitude already deemed extraordinarily intense within astrophysical settings. Accordingly, we adopt $Bm$ peaking around the order of 0.01 as a representative value for illustrating the results in the following sections.

We note that the mass parameter $m$ does not necessarily correspond to the global mass $M$ relevant for black hole thermodynamics. As shown in \cite{Podolsky:2025tle}, for non-rotating cases, the relation $M = C m$ holds. However, since $Bm$ is small in our setup, we have $C \simeq 1$ and $m$ serves as a good approximation to the black hole mass. In the following, we set $m = 1$ for simplicity.

\section{Geodesics of the KBR Exterior}\label{sec3}

In this section, we examine the geodesic motion in the KBR spacetime. 
The general Hamilton–Jacobi equation takes the form \cite{1983mtbh.book.....C, PhysRev.174.1559,Perlick:2021aok}
\begin{equation}
\frac{\partial S}{\partial \lambda} + H = 0\,,
\end{equation}  
where $S$ is the Hamilton principal function, and $\lambda$ denotes the affine parameter. For a freely moving particle subject to no external forces, the Hamiltonian $H$ is given by
\begin{equation}
H = \frac{1}{2} g^{\mu\nu} \frac{\partial S}{\partial x^\mu} \frac{\partial S}{\partial x^\nu} = -\frac{1}{2} \mu^2\,,
\end{equation}
where $\mu = 1$ for massive particles and $\mu = 0$ for massless particles.
The four-momentum of the particle is defined by $p_\mu = \partial S/\partial x^\mu$. In a stationary and axisymmetric spacetime, the existence of two Killing vectors leads to the conservation of energy and angular momentum, expressed as $\mathcal{E} = -p_t$, $\mathcal{L} = p_{\phi}$, respectively.
The Hamiltonian can be rewritten in the form $2 H = \mathcal{C} + g^{rr} p_r^2 + g^{\theta\theta} p_{\theta}^2$, where the function $\mathcal{C}$ is defined as $\mathcal{C} = g^{tt}\mathcal{E}^2 - 2g^{t\phi}\mathcal{E}\mathcal{L}  + g^{\phi\phi}\mathcal{L}^2$.
Substituting the specific form of the KBR metric from Eq.~\eqref{KBR02} into $\mathcal{C}$, we obtain
\bea
\frac{\mathcal{C}}{\Omega^2} &=& \frac{ C^2Qa\sin^2{\theta}\mathcal{E}^2 + Q \csc^2{\theta} \mathcal{L}^2 - P a^2  \mathcal{L}^2 -C^2P\left(r^2+a^2\right)^2\mathcal{E}^2 - 2aCQ  \mathcal{E}\mathcal{L} +  2aC P \left(r^2+a^2\right) \mathcal{E}\mathcal{L}}{ C^2 P Q \Sigma } \nn \\
 %&=& \frac{ 2aC \left(r^2+a^2\right)\mathcal{E}\mathcal{L} -a^2\mathcal{L}^2 - C^2 \left(r^2+a^2\right)^2\mathcal{E}^2 }{  C^2 Q \Sigma  } + \frac{  C^2a\sin^2{\theta}\mathcal{E}^2 - 2aC \mathcal{E}\mathcal{L} +  \csc^2{\theta} \mathcal{L}^2 }{  C^2 P \Sigma  } \nn \\
&=& - \frac{ \left[ C\left(r^2+a^2\right) \mathcal{E} - a \mathcal{L} \right]^2  }{  C^2 Q \Sigma  } + \frac{ \left( Ca\sin^2{\theta}\mathcal{E} - \mathcal{L} \right)^2 }{  C^2 P \Sigma \sin^2{\theta} } \,.
\eea
This leads to the Hamilton–Jacobi equation in the following form:
\begin{equation}\label{Heq}
\frac{1}{2} \mu^2 + \frac{1}{2} \frac{\Omega^2}{\Sigma} \left[ Q \left(\frac{\partial S}{\partial r}\right)^2 - \frac{\left[(a^2 + r^2)\mathcal{E} - a C^{-1}\mathcal{L}\right]^2}{Q} + P \left(\frac{\partial S}{\partial \theta}\right)^2 + \frac{\left[a \sin^2{\theta} \, \mathcal{E} - C^{-1}\mathcal{L}\right]^2}{P \sin^2{\theta}} \right] = 0.
\end{equation}
Although the terms within the square brackets are separable in $r$ and $\theta$, the prefactor $\Sigma^{-1}\Omega^2$, which depends on both variables, generally prevents full separation of variables in the Hamilton–Jacobi equation for timelike particles ($\mu \neq 0$). Consequently, the geodesic equations are, in general, not separable for massive particles. However, for null geodesics 
($\mu = 0$), the equation becomes separable. In this case, the Hamilton principal function can be written as  $S = S_r(r) + S_{\theta}(\theta) -\mathcal{E} t +\mathcal{L} \phi$. This separability arises from the fact that the conformal factor does not affect the trajectories of null geodesics.
Finally, we note that the parameter $C$ can be absorbed into the angular momentum by the rescaling $\mathcal{L} \rightarrow C\mathcal{L}$ in Eq.~\eqref{Heq}. Without loss of generality, we set $C = 1$ for the remainder of this analysis.

\subsection{Null geodesics and photon region}

In this subsection, we focus on null geodesics. In this case, the radial and angular components of the four-momentum are given by $p_r = \partial_r S_r(r)$ and $p_{\theta} =\partial_{\theta} S_{\theta}(\theta)$, respectively. Setting $\mu = 0$ in Eq.~\eqref{Heq}, the equations of motion in the $r$ and $\theta$ directions become
\begin{equation}
Q p_r^2 - \frac{\left[(a^2 + r^2)\mathcal{E} - a\mathcal{L}\right]^2}{Q} = -\kappa\,,\quad
P p_\theta^2 + \frac{\left(a \sin^2{\theta} \, \mathcal{E} - \mathcal{L} \right)^2}{P \sin^2{\theta}} = \kappa\,,
\end{equation}
where $\kappa$  is a separation constant arising from the separability of the Hamilton–Jacobi equation. This constant is independent of the particle’s mass, energy, and angular momentum. It is convenient to define a shifted conserved quantity,
\begin{equation}
\mathcal{Q} = \kappa - (\mathcal{L} - a\mathcal{E})^2\,,
\end{equation}  
which allows the equations of motion to be recast in the form 
\begin{equation}
Q^2 p_r^2 = R(r)\,,\quad P^2 p_\theta^2 = \Theta(\theta)\,,
\end{equation}  
where the functions $R$ and $\Theta$ represent the effective radial and angular potentials, respectively:
\begin{equation}
\begin{aligned}
R(r) &= \left[(a^2 + r^2)\mathcal{E} - a\mathcal{L}\right]^2 - \Delta\left[\mathcal{Q} + (\mathcal{L} - a\mathcal{E})^2\right]- B^2 r^2 \Delta\left[\mathcal{Q} + (\mathcal{L} - a\mathcal{E})^2\right]\,, \\
\Theta(\theta) &= \mathcal{Q} - \left(\mathcal{L}^2 \csc^2\theta - a^2 \mathcal{E}^2\right)\cos^2\theta + B^2\left(\frac{I_2}{I_1^2} - a^2\right)\left[(\mathcal{L} - a\mathcal{E})^2 + \mathcal{Q}\right] \cos^2\theta\,.
\end{aligned}
\end{equation}  
The full set of equations governing null geodesics is then given by
\begin{equation}\label{geq}
\begin{aligned}
\frac{\Sigma}{\Omega^2}\frac{\mathrm{d}t}{\mathrm{d}\lambda} &= \frac{r^2 + a^2}{Q}\left[\mathcal{E}(r^2 + a^2) - a\mathcal{L}\right] + \frac{a}{P}(\mathcal{L} - a\mathcal{E} \sin^2\theta)\,, \\
\frac{\Sigma}{\Omega^2}\frac{\mathrm{d}r}{\mathrm{d}\lambda} &= \pm_r \sqrt{R(r)}\,, \\
\frac{\Sigma}{\Omega^2}\frac{\mathrm{d}\theta}{\mathrm{d}\lambda} &= \pm_\theta \sqrt{\Theta(\theta)}\,, \\
\frac{\Sigma}{\Omega^2}\frac{\mathrm{d}\phi}{\mathrm{d}\lambda} &= \frac{a}{Q}\left[\mathcal{E}(r^2 + a^2) - a\mathcal{L}\right] + \frac{\mathcal{L}}{P \sin^2\theta} - \frac{a\mathcal{E}}{P}\,.
\end{aligned}
\end{equation}  
These equations reduce to those of the Kerr spacetime in the limit $B \to 0$. Since rescaling the energy $\mathcal{E}$ does not affect the trajectory of photons, it is convenient to introduce the dimensionless impact parameters
\begin{equation}
\xi = \frac{\mathcal{L}}{\mathcal{E}}\,, \quad \eta = \frac{\mathcal{Q}}{\mathcal{E}^2}\,.
\end{equation}
To compute null geodesics, we first solve the lensing equation, which is obtained from the ratio of the radial and angular components in Eq.~\eqref{geq}, yielding a relation between the variations in $r$ and $\theta$. Once this relation is established, the equations for $t$ and $\phi$ can be integrated to determine the corresponding changes in coordinate time and azimuthal angle.

A preliminary step in analyzing null geodesics involves studying the effective potentials, particularly their critical behavior. Introducing the variable $u = \cos^2{\theta}$, the angular potential can be recast as
\begin{equation}
\Theta(u) = \frac{1}{1-u} \left(\eta + A_1 u - A_2 u^2\right) = \frac{A_2}{1-u}\left( u_+ - u\right) \left( u - u_- \right)  \,,
\end{equation}
where
\begin{equation}
\begin{aligned}
A_1 &= \left[a^2  - \xi^2 - \eta - B^2 \left(a^2 - \frac{I_2}{I_1^2} \right)\left[(a - \xi)^2  +\eta\right]\right]\,, \\
A_2 &= \left[a^2 - B^2 \left(a^2 - \frac{I_2}{I_1^2} \right)\left[(a - \xi)^2  +\eta\right] \right] \,,\quad u_{\pm} = \frac{A_1 \pm \sqrt{A_1^2 + 4\eta A_2}}{2A_2} \,.
\end{aligned}
\end{equation}
For small values of $B$, and assuming $a \neq 0$, we have $A_2 > 0$, implying that the qualitative behavior of motion in the $\theta$ direction remains similar to that in the Kerr family spacetimes \cite{Gralla:2019drh,Hou:2022gge}. Since $\Theta|_{u \to 1} \to -\xi^2\left(1-u\right)^{-1} $, only photons with $\xi = 0$ can reach the polar axis. When $\eta \geq 0$, one finds $u_- < 0$, meaning that $u\in[0,u_+]$, and photons can cross the equatorial plane ($u = 0$), oscillating within the range $\theta \in [\theta_+ , \pi - \theta_+]$. Conversely, for $\eta < 0$, we have $u_- > 0$, and photons are confined to the regions $\theta \in [\theta_- , \theta_+]$ or $ [\pi - \theta_+ , \pi - \theta_-]$, where $\theta_{\pm} = \arccos{\sqrt{u_{\pm}}}$.

The radial potential $R(r)$ is an eighth-degree polynomial in $r$, making a complete analytical treatment intractable. However, the properties of spherical photon orbits are governed solely by the local extrema of $R(r)$, specifically the points where $R = \partial_r R = 0$. These conditions yield constraints on the impact parameters:
\begin{equation}\label{xeeq}
\begin{aligned}
\xi_p = \frac{r_p^2 + a^2}{a} - \frac{4 r_p Q(r_p)}{a Q'(r_p)}\,, \quad
\eta_p = \frac{16 r_p^2 Q(r_p)}{Q'(r_p)^2} - \left( \frac{r_p^2}{a} - \frac{4 r_p Q(r_p)}{a Q'(r_p)} \right)^2\,.
\end{aligned}
\end{equation}
Here, the subscript ``p'' denotes evaluation at the radius of the spherical photon orbit, $r = r_p$. As discussed earlier, equator-crossing null geodesics correspond to $\eta \geq 0$. 
Combining this condition with Eq.\eqref{xeeq}, the photon region is bounded by the inequality
\begin{equation}\label{rneq}
\left(4 r_p Q(r_p) - r_p^2 Q'(r_p)\right)^2 \leq 16 a^2 r_p^2 Q(r_p)\,,
\end{equation}
%\begin{equation}\label{rneq}
%\left(4 r_p Q - \Sigma Q'\right)^2 \leq 16 a^2 r_p^2 Q P \sin^2\theta\,,
%\end{equation}
which typically defines the radial extent of the photon region as $r_p \in [r_{p-},\, r_{p+}]$. At the boundaries $r = r_{p\pm}$, we have $\eta_p = 0$, and the corresponding photons follow circular orbits confined to the equatorial plane. Solving Eq.\eqref{rneq} as an equality yields the explicit expressions for $r_{p\pm}$. Assuming a small magnetic field parameter $B$, the radial boundaries of the photon region can be expanded perturbatively in $B$ as
\begin{equation}
r_{p\pm} = r^K_{p\pm} + \delta r_{p\pm}\,,
\end{equation}  
where  
\begin{equation}
r^K_{p\pm} = 2\left(1 + \cos\varphi_\pm\right)\,, \quad \varphi_\pm = \frac{2}{3} \cos^{-1}(\pm a)\,,
\end{equation}  
are the corresponding boundaries in the Kerr spacetime, and
\begin{equation}
\delta r_{p\pm} = B^2 \frac{ (\cos^2\varphi_\pm + 1) \left[ 3a^4 + 48a^2 + 7(5a^2 + 4)\cos\varphi_\pm + (11a^2 + 26)\cos 2\varphi_\pm + 2\cos 4\varphi_\pm \right]}{3\left[a^2 + \cos\varphi_\pm + \cos 2\varphi_\pm\right]}\,,
\end{equation}  
represents the first-order correction to the photon region due to the magnetic field in the KBR spacetime.

\subsection{Circular timelike geodesics and the ISCO}

In general, the equations of motion for timelike particles are not separable in the same manner as those for null geodesics. In this subsection, we briefly analyze the circular motion of timelike geodesics confined to the equatorial plane. Within this plane, $\theta = \pi/2$ and $p_{\theta} = 0$,  which simplifies the equations of motion. Under these conditions, the radial equation of motion can be recast in the form
%\begin{equation}
%1 + \frac{\left(1 + B^2 r^2 \right)}{r^2} \left[ Q p_r^2 - \frac{\left[(a^2 + r^2)\mathcal{E} - a\mathcal{L} \right]^2}{Q} + (a\mathcal{E} - \mathcal{L})^2 \right] = 0\,.
%\end{equation}  
$p_r^2+ V(\mathcal{E},\mathcal{L},r)=0$,
where the effective potential $V$ is given by 
\begin{equation}
    V(\mathcal{E},\mathcal{L},r)=\frac{r^2}{Q\left(1 + B^2 r^2 \right)}-\frac{\left[(a^2+r^2)\mathcal{E}-a\mathcal{L}\right]^2}{Q^2}+\frac{(a\mathcal{E}-\mathcal{L})^2}{Q}\,.
\end{equation}
Circular orbits correspond to extrema of the effective potential and are determined by the conditions
$V = \partial_r V = 0$, which constrain the energy $\mathcal{E}$ and angular momentum $\mathcal{L}$ in terms of the orbital radius. 
The angular velocity $\Omega_c$ of a particle in a circular orbit can be obtained from the condition $\partial_r g_{tt} + 2 \partial_r g_{t\phi}\Omega_c + \partial_r g_{\phi\phi}\Omega_c^2=0$. For a weak magnetic field $B \ll 1$, the angular velocity can be expanded as
\begin{equation}
\Omega_c = \frac{1}{a \pm r^{3/2}} + B^2 \Omega_{(B)} \,,
\end{equation}
where the first term corresponds to the Kerr angular velocity, and the second term $\Omega_{(B)} $ represents the leading-order correction due to the magnetic field:
\begin{equation}
\Omega_{(B)} = \frac{ r^{3/2}\left( a^2\pm4 a r^{3/2}-4 r^2\right) }{4\left(a\pm r^{3/2}\right)^2}\,.
\end{equation}
Not all circular orbits are stable. In particular, orbits sufficiently close to the horizon become unstable under radial perturbations, causing particles to plunge into the black hole. The innermost stable circular orbit (ISCO) is defined by the marginal stability condition:
\begin{equation}
V = 0, \quad \partial_r V = 0, \quad \partial_r^2 V = 0\,.
\end{equation}
Assuming again that $B \ll 1$, the ISCO radius can be expressed perturbatively as $r_{\text{isco}} = r_0 + B^2 r_{(B)}$, where $r_0$ is the ISCO radius in the Kerr spacetime, given by
\begin{equation}
\begin{aligned}
Z_1 &= 1 + (1 - a^2)^{1/3} \left[(1 + a)^{1/3} + (1 - a)^{1/3} \right]\,, \\
Z_2 &= \sqrt{3 a^2 + Z_1^2}\,, \\
r_0 &= 3 + Z_2 \mp \sqrt{(3 - Z_1)(Z_1 + 2 Z_2 + 3)}\,,
\end{aligned}
\end{equation}
with the upper (lower) sign corresponding to corotating (counterrotating) orbits.
The correction $r_{(B)}$ encodes the leading-order effect of the magnetic field on the ISCO radius and takes the form
\begin{equation}
r_{(B)} =  \bigg[
21 a^6 r_0
+ r_0^2 (27 a^6 - 84 a^4)
+ r_0^3 (60 a^2 - 82 a^4)
+ r_0^4 (6 a^4 + 236 a^2)
\end{equation}
\begin{equation*}
{} + r_0^5 (-103 a^2 - 240)
+ r_0^6 (228 - 17 a^2)
- 56 r_0^7
+ 4 r_0^8
\bigg]\,.
\end{equation*}

\section{Shadows of the KBR black hole}\label{sec4}

\subsection{Set up of the observer}
In this section, we explore the properties of the shadow cast by the KBR black hole. Firstly, it is essential to construct the observer's local orthonormal frame. Here, we adopt the widely used Zero Angular Momentum Observer (ZAMO) frame \cite{Bardeen:1973tla, Frolov:1998wf, Johannsen:2013vgc}, whose associated tetrad vectors are given by:
\bea\label{zamocoorbasis}
\hat{e}_{(0)} = \zeta\, \partial_t + \Gamma\, \partial_\phi\,, \quad
\hat{e}_{(1)} = -\frac{\partial_r}{\sqrt{g_{r r}}}\,, \quad
\hat{e}_{(2)} = \frac{\partial_\theta}{\sqrt{g_{\theta \theta}}}\,, \quad
\hat{e}_{(3)} = -\frac{\partial_\phi}{\sqrt{g_{\phi \phi}}}\,,
\eea
where
\begin{equation}
    \begin{split}
        \zeta = \frac{g_{\phi \phi}}{\sqrt{g_{\phi \phi} \left(g_{\phi t}^2 - g_{\phi \phi} g_{t t}\right)}}\,, \qquad 
        \Gamma = \frac{-g_{\phi t}}{\sqrt{g_{\phi \phi} \left(g_{\phi t}^2 - g_{\phi \phi} g_{t t}\right)}}\,.
    \end{split}
\end{equation}
The photon's four-momentum in the ZAMO frame is expressed as $p_{(a)} = p_\nu e_{(a)}^\nu$. 
%By employing celestial coordinates, we project the arrivaling photon's momentum onto the observational plane. A Cartesian coordinate system is established on the observational screen,  with the $x$-axis -axis aligned with the basis vector $e_{(3)}$, the $y$-axis with $e_{(2)}$, and the origin shifted by one unit along the negative $e_{(1)}$-axis in the ZAMO frame.
To facilitate the reconstruction of the black hole shadow on the observational screen, we employ the stereographic projection method as outlined in \cite{Hu:2020usx}. The celestial coordinates $\Theta$ and $\Psi$, which characterize the angular position of incoming photons, are determined by the tetrad components of the photon's four-momentum $p_{(a)}$:
\bea  
\cos \Theta = \frac{p^{(1)}}{p^{(0)}}, \quad \tan \Psi = \frac{p^{(3)}}{p^{(2)}}\,.  
\eea  
It is worth noting that, under the definition, the range of $\Psi$ spans from $0$ to $\pi$. The corresponding Cartesian coordinates $(x, y)$ on the image plane are then obtained via the stereographic projection formulas:
\bea  
x = -2 \tan \left( \frac{\Theta}{2} \right) \sin \Psi\,, \quad y = -2 \tan \left( \frac{\Theta}{2} \right) \cos \Psi \,. 
\eea  
This projection maps the celestial sphere onto a two-dimensional observational screen, enabling a faithful visualization of the black hole shadow.

\begin{figure}[t]
    \centering
    \includegraphics[width=4.3in]{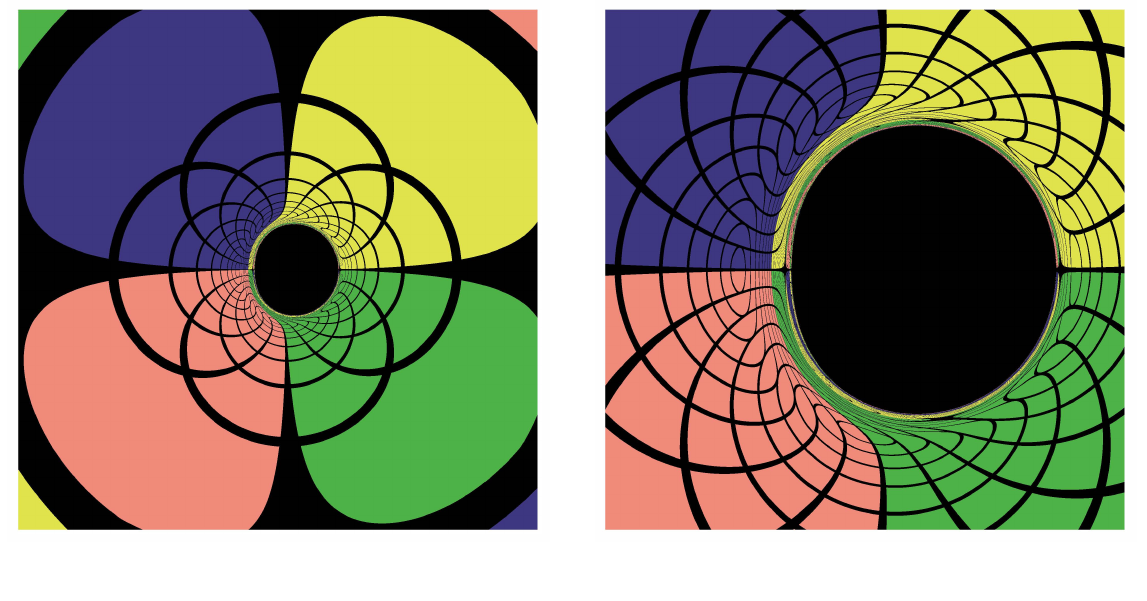}\includegraphics[width=2.2in]{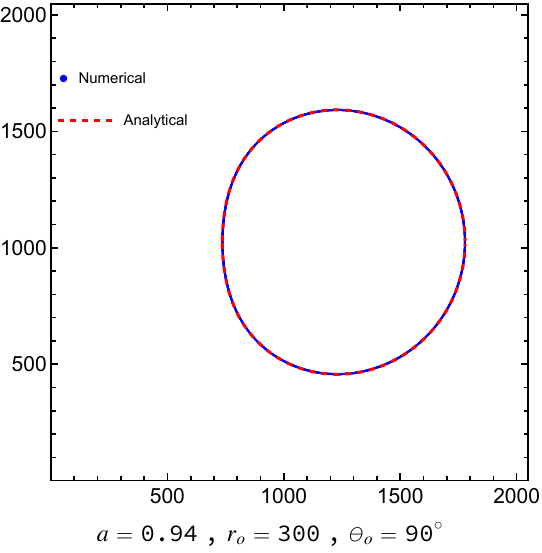}
    \caption{Shadows of KBR black holes for $B=0$ (\textbf{Left}) and $B=0.01$ (\textbf{Middle}), with fixed parameters $a=0.94\,,r_o=300\,,\theta_o=90^\circ\,$. \textbf{Right:} Critical curves obtained via numerical ray tracing (blue curve) and the analytical method (red dashed curve). The parameters are fixed to $a=0.94\,,B=0.01\,,r_o=300\,,\theta_o=90^\circ\,$.}
    \label{fig:lensingshadow}
\end{figure}

Figure~\ref{fig:lensingshadow} illustrates the black hole shadow illuminated by a celestial sphere divided into four colored quadrants—a common setup adopted in the literature (see, e.g., \cite{Cunha:2018acu,Hu:2020usx,Chen:2022scf}). In the left panel, one can clearly observe how strong gravitational lensing distorts the background celestial sphere, producing multiple higher-order images near the central dark region, known as the shadow. The boundary of the shadow is defined by the critical curve (the right panel), which corresponds to the projection of unstable spherical photon orbits within the photon region. The four-momentum of photons arriving at the observer from this curve is characterized by the impact parameters given in Eq.~\eqref{xeeq}.

The right panel of Figure~\ref{fig:lensingshadow} presents a comparison between the analytically derived critical curve and the numerically computed shadow contour on the celestial sphere. As expected, they exhibit close agreement.

\subsection{The critical curve}

\begin{figure}[t]
    \centering
    \includegraphics[width=4.5in]{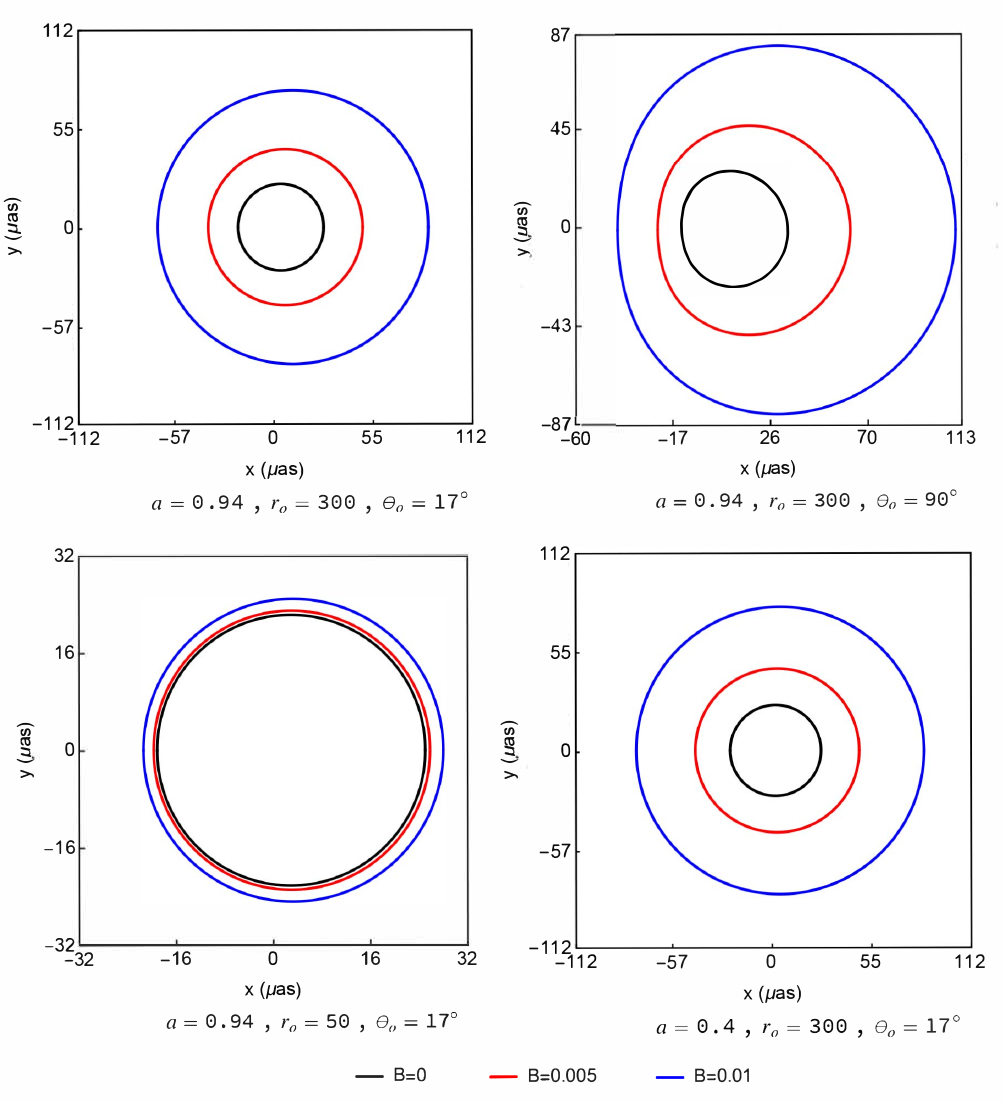}
    \caption{Variations of the critical curve under different black hole parameters and observer locations.}
    \label{fig1}
\end{figure}

In this subsection, we present the critical curve structures in the KBR spacetime. Figure~\ref{fig1} illustrates how the critical curves vary with different black hole parameters and observer positions. 
The top-left panel shows the case of a relatively high spin parameter, $a = 0.94$, with the observer located at a small inclination angle $\theta_o = 17^{\circ}$. In this case, the critical curve is nearly circular. Notably, the size of the shadow is sensitive to the magnetic field strength $B$: increasing $B$ enlarges the critical curve, while its circularity remains largely unaffected.
A similar trend is observed in the top-right panel, where the observer is placed at the equatorial plane ($\theta_o = 90^{\circ}$). However, due to the black hole’s rotation, the critical curve exhibits a characteristic “D”-shaped distortion.

The bottom-left panel displays the results for an observer located closer to the black hole, at $r_o = 50$. In this case, the influence of $B$ on the shadow size becomes less pronounced. This behavior can be understood by introducing a characteristic radius $R_B = 1/B$. When $B$ is small, we have approximately $aB \leq B \ll 1$, implying $R_B \gg 1$. 
In the asymptotically defined near zone, $r_+ \leq r \ll R_B$, the term proportional to $B^2 r^2$ remains subdominant, and the spacetime effectively reduces to the Kerr geometry, as can be directly verified from the metric components in Eq.~\eqref{KBR}. Conversely, in the far zone, $r \gg R_B$,  the spacetime transitions to a regime dominated by the Bertotti–Robinson geometry, characterized by an $AdS_2 \times S^2$ symmetry. In this region, the asymptotic AdS-like structure induced by the magnetic field significantly modifies photon trajectories, leading to substantial deviations in the black hole shadow compared to the Kerr case. 

Applying this analysis to the case shown in Figure~\ref{fig1}, the characteristic radius takes the values of $100$ for $B = 0.01$ and $200$ for $B = 0.005$. Accordingly, an observer located at $r_o = 300$ effectively resides in the far zone, where the asymptotic deviations from the Kerr geometry become significant. This observer is thus capable of detecting the modifications to the black hole shadow induced by the magnetic field. In contrast, an observer situated at $r_o = 50$ effectively lies well within the near zone, where the spacetime remains effectively Kerr-like and the asymptotically AdS-like structure is not observable.

\subsection{Derivations from Kerr}

In this subsection, we proceed to study the derivation of critical curves in KBR spacetime, comparing them with those of the Kerr black hole. To provide a quantitative characterization of this discrepancy, and in accordance with our earlier work \cite{Zhong:2021mty}, we introduce the coordinates $(x_c, y_c)$ on the observation screen, defined as  
\[
x_c = \frac{x_{\min} + x_{\max}}{2}\,, \quad y_c = \frac{y_{\min} + y_{\max}}{2}\,,
\]  
where $x_{\min}$, $x_{\max}$, $y_{\min}$, and $y_{\max}$ denote the minimal and maximal horizontal and vertical coordinates of the critical curve, respectively. These coordinates correspond to the geometric centre of the projected region. Given that for the same pair $(\xi, \eta)$, it is permissible for $p_\theta$ to assume either a positive or negative sign—as allowed by Eq.~(\ref{geq})—we consequently always have $y_{\min} + y_{\max} = 0=y_c$. We introduce polar coordinates $(\rho, \psi)$ centered at this point, where  
$\rho = \sqrt{(x - x_c)^2 + y^2}$. The averaged radius of the critical curve is then defined as  
\[
\bar{\rho} = \int_0^{2\pi} \frac{\rho(\psi)}{2\pi} \, d\psi\,.
\]
To quantify the deviation from the Kerr case, we define the dimensionless parameter  
\[
\sigma = \frac{\bar{\rho}}{\bar{\rho}_0} - 1\,,
\]
where the subscript “0” refers to the corresponding quantity in the Kerr spacetime. Such dimensionless parameter characterizes the deformation of the KBR black hole shadow relative to that of a Kerr black hole.

\begin{figure}[htbp]
    \centering
    \includegraphics[width=7in]{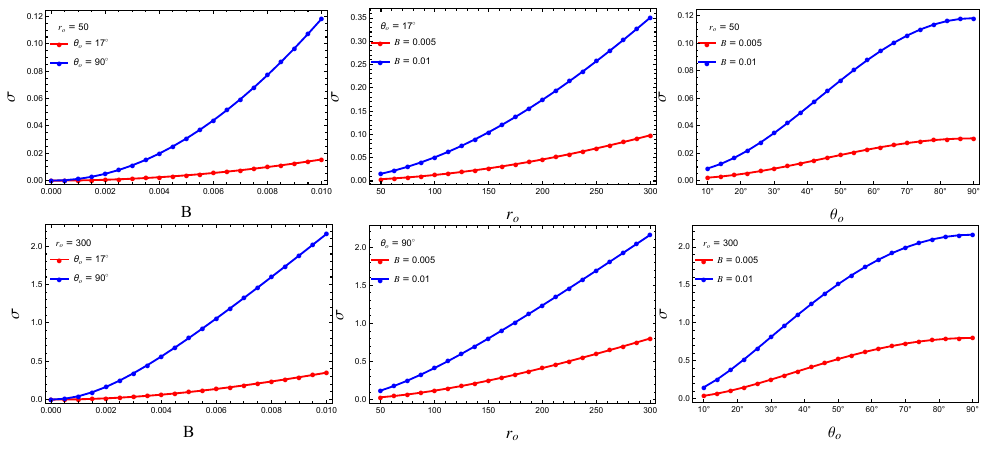}
    \caption{Variation of the dimensionless parameter $\sigma=\bar{\rho}/\bar{\rho}_0-1$ as a function of the magnetic field strength $B$ (left column), observer distance $r_o$ (middle column) and inclination angle $\theta_o$ (right column).}
    \label{fig2}
\end{figure}

Figure~\ref{fig2} illustrates the dependence of the deformation parameter $\sigma$ on the magnetic field strength $B$, observer distance $r_o$, and inclination angle $\theta_o$.  
The left column displays $\sigma$ as a function of $B$ for two fixed inclination angles, $\theta_o = 17^\circ$ and $\theta_o = 90^\circ$, with the top and bottom panels corresponding to observer distances $r_o = 50$ and $r_o = 300$, respectively. 
Evidently, $\sigma$ increases with the magnetic field strength. This behavior is expected, as a stronger magnetic field amplifies the deviations of the KBR black hole from the standard Kerr spacetime. As $B$ increases from zero, $\sigma$ initially grows slowly, and subsequently enters an approximately linear growth regime.

Comparing the upper and lower panels, when $B$ remains relatively small, we note that, for the same value of $B$, the case with $r_o = 50$ exhibits a closer resemblance to the Kerr black hole than the case with $r_o = 300$. This can be attributed to the existence of the characteristic radius $R_B = 1/B$, which defines a critical magnetic scale. Compared to the observer at $r_o = 50$, the observer located at $r_o = 300$ lies closer to the asymptotic far zone $r \gg R_B$, 
where the AdS-like asymptotic structure induced by the magnetic field significantly modifies photon trajectories, resulting in pronounced deviations in the black hole shadow relative to the Kerr scenario. At this stage, the relationship between $\sigma$ and $B$ becomes approximately linear. Moreover, the shadow deformation is significantly more pronounced at $\theta_o = 90^\circ$ than at $\theta_o = 17^\circ$, with the disparity increasing as $B$ grows, indicating that the influence of the inclination angle on shadow distortion is enhanced in regimes with stronger magnetic fields.

The middle column illustrates the dependence of $\sigma$ on the observer distance $r_o$ for fixed magnetic field strengths, $B = 0.005$ and $B = 0.01$. The top and bottom panels correspond to inclination angles $\theta_o = 17^\circ$ and $\theta_o = 90^\circ$, respectively. As shown, $\sigma$ increases significantly with increasing $r_o$, indicating that shadow deformation becomes more pronounced as the observer moves farther away. Furthermore, the difference in $\sigma$ between the two magnetic field strengths becomes increasingly evident at larger observer distances, suggesting an enhanced sensitivity to $B$ in the far-field regime.

The right column depicts the variation of $\sigma$ with the inclination angle, for fixed magnetic field strengths $B = 0.005$ and $B = 0.01$. The top and bottom panels correspond to observer distances $r_o = 300$ and $r_o = 50$, respectively. In both cases, $\sigma$ increases with $\theta_o$, gradually saturating and reaching its maximum as $\theta_o$ approaches $90^\circ$.

Across all panels, the dimensionless deformation parameter $\sigma$ increases with both the magnetic field strength and the inclination angle, indicating more significant deviations from the Kerr shadow. Additionally, $\sigma$ exhibits clear growth with increasing observer distance, suggesting that the deformation becomes more pronounced for distant observers.

\section{Summary}\label{sec5}

In this work, we investigated the geodesics and the shadow cast by a black hole in the spacetime described by a newly proposed solution\cite{Podolsky:2025tle} for a Kerr black hole immersed in a uniform magnetic field. 
Through the Hamilton–Jacobi equation, we found that for null geodesics, the method of separation of variables can be employed to derive an analytic expression for the four-momentum. Subsequently, we analysed and discussed the radial and angular potentials governing null geodesics. In the regime of a small magnetic field $B$, we obtained an approximate analytic expression for the photon ring. In contrast, for timelike geodesics, the separation of variables proves inapplicable; thus, we focused on equatorial motion. Under the assumption of a weak magnetic field, we derived approximate expressions for the angular velocity of circular orbits and the ISCO radius.

Subsequently, we conducted a systematic investigation into the shadow structure of the KBR black hole. By employing both numerical and analytical methods, we performed a cross-validation of our results and found them to be consistent. We not only carried out a qualitative analysis of how the magnetic field strength, observer distance, inclination angle, and black hole spin influence the KBR black hole shadow, but also provided a quantitative assessment of the deviations between the KBR and Kerr shadows as these parameters vary.

We observed that as the observation angle approaches $90^\circ$, the degree of deviation becomes more pronounced. While variations in the observer’s distance and magnetic field strength generally lead to increased deviations, certain subtle distinctions arise when the magnetic field is relatively weak or the observer is positioned closer to the black hole. By introducing a characteristic radius and defining the asymptotically defined near zone, $r_+ \leq r \ll R_B$, and the far zone, $r \gg R_B$, we found that the spacetime within the near zone is approximately asymptotically flat, thereby yielding shadow structures closely resembling those of Kerr black holes. In contrast, the far zone exhibits an asymptotically AdS-like structure induced by the magnetic field, resulting in more significant deviations. This classification offers a self-consistent explanation for the observed differences.

We conclude this paper with several prospects for future exploration. Firstly, upcoming studies could further investigate the images of KBR black holes with accretion disks serving as the light source \cite{Hou:2022eev, Zhang:2024lsf}, focusing on how the inner shadow structure diverges from that of the Kerr case. Secondly, given that the KBR spacetime self-consistently incorporates a magnetic field, it presents a natural avenue for exploring polarization images induced by the magnetic field, thereby deepening our understanding of such phenomena. Moreover, drawing comparisons between the findings in the KBR spacetime and those in the Kerr–Melvin spacetime would be a highly worthwhile endeavor.

\section*{Acknowledgments}
We are grateful to Jinchi Liu for his valuable support. The work is partly supported by NSFC Grant No. 12205013, 12275004 and 12588101. Y. H. acknowledges support from the NSFC Grant No. 12547123. M. G. is also supported by Open Fund of Key Laboratory of Multiscale Spin Physics (Ministry of Education), Beijing Normal University and he is also supported by the BNU Tang Scholar.

\appendix
\label{app}

\bibliographystyle{utphys}
\bibliography{reference}

\end{document}